\documentclass[aps,prl,twocolumn,showpacs,superscriptaddress,floatfix]{revtex4-1}

\pdfoutput=1

\usepackage{amsmath,amssymb}
\usepackage{graphicx}
\usepackage{color}
\usepackage{rotating}
\usepackage{bm}
\usepackage{setspace}
\usepackage{hyperref}

\hypersetup{
colorlinks=true,
urlcolor= blue,
citecolor=blue,
linkcolor= blue,
bookmarks=true,
bookmarksopen=false,
}

\renewcommand{\vec}[1]{\bm{#1}}

\begin{document}


\title{Revealing the Empty-State Electronic Structure of Single-Unit-Cell FeSe/SrTiO$_{3}$}

\author{Dennis Huang}
\affiliation{Department of Physics, Harvard University, Cambridge, MA 02138, USA}
\author{Can-Li Song}
\altaffiliation[]{Present address: State Key Laboratory of Low-Dimensional Quantum Physics, Department of Physics, Tsinghua University, Beijing 100084, China}
\affiliation{Department of Physics, Harvard University, Cambridge, MA 02138, USA}
\author{Tatiana A. Webb}
\affiliation{Department of Physics, Harvard University, Cambridge, MA 02138, USA}
\affiliation{Department of Physics \& Astronomy, University of British Columbia, Vancouver, British Columbia V6T 1Z1, Canada}
\author{Shiang Fang}
\affiliation{Department of Physics, Harvard University, Cambridge, MA 02138, USA}
\author{\\Cui-Zu Chang}
\affiliation{Francis Bitter Magnet Laboratory, Massachusetts Institute of Technology, Cambridge, MA 02139, USA}
\author{Jagadeesh S. Moodera}
\affiliation{Francis Bitter Magnet Laboratory, Massachusetts Institute of Technology, Cambridge, MA 02139, USA}
\affiliation{Department of Physics, Massachusetts Institute of Technology, Cambridge, MA 02139, USA}
\author{Efthimios Kaxiras}
\affiliation{Department of Physics, Harvard University, Cambridge, MA 02138, USA}
\affiliation{School of Engineering and Applied Sciences, Harvard University, Cambridge, MA 02138, USA}
\author{Jennifer E. Hoffman}
\email[]{jhoffman@physics.harvard.edu}
\affiliation{Department of Physics, Harvard University, Cambridge, MA 02138, USA}

\date{\today}


\begin{abstract}
We use scanning tunneling spectroscopy to investigate the filled and empty electronic states of superconducting single-unit-cell FeSe deposited on SrTiO$_3$(001). We map the momentum-space band structure by combining quasiparticle interference imaging with decay length spectroscopy. In addition to quantifying the filled-state bands, we discover a $\Gamma$-centered electron pocket 75 meV above the Fermi energy. Our density functional theory calculations show the orbital nature of empty states at $\Gamma$ and explain how the Se height is a key tuning parameter of their energies, with broad implications for electronic properties.
\end{abstract}


\maketitle

The extraordinary potential of interface engineering to generate novel electronic properties is exemplified by a single unit cell (1UC) of FeSe deposited on SrTiO$_3$ \cite{Wang_CPL_2012}, which exhibits an order-of-magnitude increase in its superconducting transition temperature ($T_c$ up to 110 K \cite{Ge_NatMat_2014}) compared to bulk FeSe ($T_c$ = 9.4 K \cite{Song_JKPS_2011}). Not only does this finding elevate the $T_c$ of iron-based superconductors (Fe-SCs) above the liquid nitrogen temperature, it also opens the door to designing Fe-SC/oxide heterostructures with novel phases and yet higher $T_c$. A key to understanding and realizing these phases is a complete measurement of the electronic structure of filled and empty states.

Electronic band structure is pivotal in determining the pairing symmetry of Fe-SCs. The generic Fermi surface (FS) of Fe-SCs consists of electron pockets at the Brillouin zone (BZ) corner M and hole pockets at the zone center $\Gamma$ \cite{Mazin_Nat_2010}. A prevalent spin-fluctuation  model suggests that repulsive antiferromagnetic excitations of wave vector ($\pi$, $\pi$) can give rise to pairing between the electron and hole pockets if the order parameter reverses sign, resulting in $s_{+-}$ superconductivity \cite{Mazin_PRL_2008, Kuroki_PRL_2008}. However, in 1UC FeSe/SrTiO$_3$, the $\Gamma$ hole pocket sinks entirely below the Fermi energy ($E_F$) due to electron doping \cite{Liu_NatComm_2012}. This challenges the $s_{+-}$ picture; nevertheless, functional renormalization group (FRG) calculations have shown that electronic bands lying within the spin fluctuation energy scale below $E_F$ can still influence the pairing channel. In fact, the energy of the sunken $\Gamma$ hole pocket is predicted to toggle the relative stability between sign-preserving $s_{++}$ and sign-changing $d$ pairing symmetries \cite{Wang_EPL_2011, Xiang_PRB_2012}.

A natural question is whether low-lying bands above $E_F$ can similarly renormalize the effective interaction. In general, the landscape of empty states in Fe-SCs remains largely unexplored by experiment. A full band structure mapping is particularly crucial in 1UC FeSe/SrTiO$_3$, where in addition to the usual Coulomb repulsion and spin fluctuations, even higher energy phonon modes may be at play \cite{Xiang_PRB_2012, Lee_Nat_2014, Coh_arXiv_2014}, and the magnitudes of their energy scales relative to the near-$E_F$ bands determine the superconducting ground state.

\begin{figure}[b]
\includegraphics[scale=1]{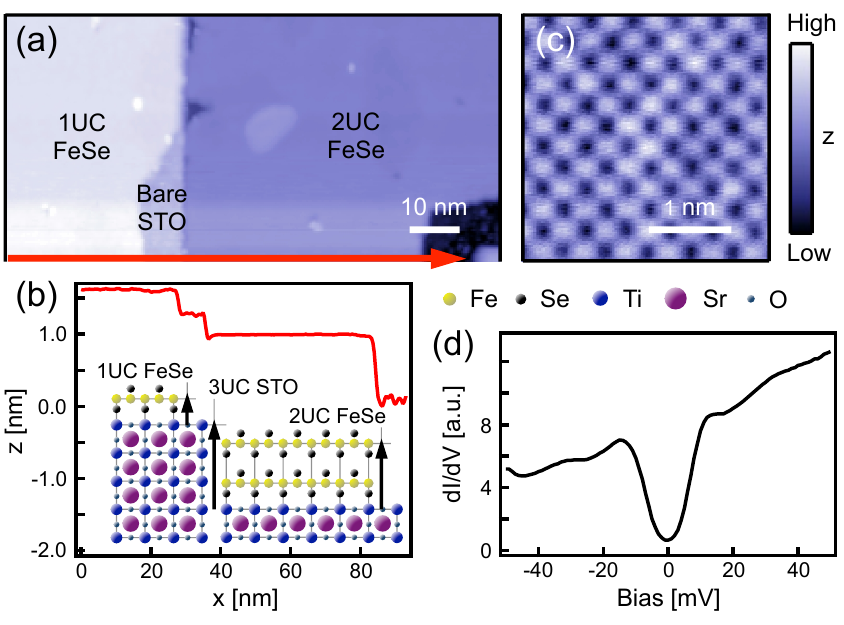}
\caption{(color online) (a) Typical topography of $\textit{in-situ}$-grown FeSe/SrTiO$_3$. Setpoint: 4 V, 5 pA. (b) Line cut along the arrow in (a). The inset illustrates the underlying crystal structure. (c) Atomically-resolved topography of single-unit-cell (1UC) FeSe/SrTiO$_3$. Setpoint: 50 mV, 250 pA. (d) $dI/dV$ spectrum of 1UC FeSe/SrTiO$_3$, T = 4.3 K. Bias oscillation $V_{\textnormal{rms}}$ = 0.7 mV.}
\label{Fig1}
\end{figure}

\begin{figure*}
\includegraphics[scale=1]{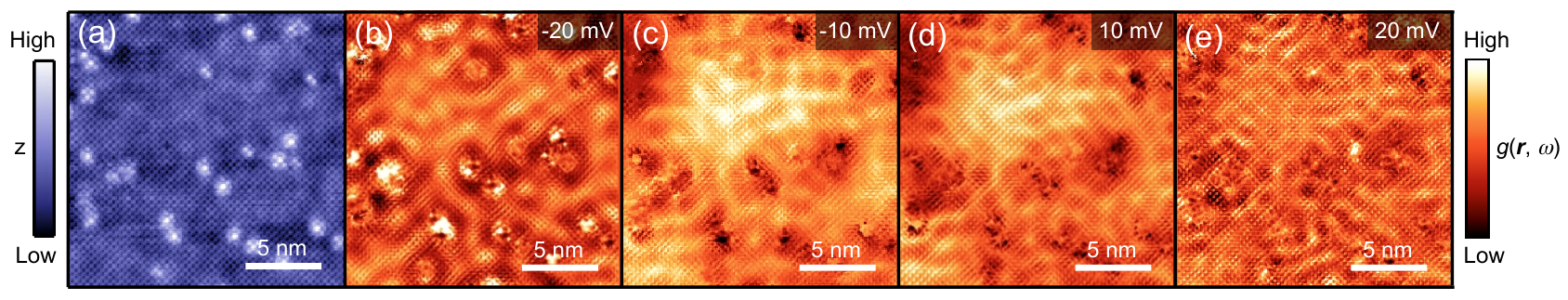}
\caption{(color online) Quasiparticle interference imaging, real space. (a) Topography (setpoint: 50 mV, 500 pA) and (b-e) conductance maps $g(\textbf{r}, \omega)$ (setpoint: 100 M$\Omega$, $V_{\textnormal{rms}}$ = 1.4 mV) of a 20 nm x 20 nm field of view with as-grown defects. Images were drift-corrected following Ref.~\cite{Lawler_Nat_2010}.}
\label{Fig2}
\end{figure*}

Here we map the multiband electronic structure of 1UC FeSe/SrTiO$_3$ by two complimentary scanning tunneling microscopy (STM) techniques: (1) quasiparticle interference (QPI) imaging \cite{Hoffman_Science_2002} and (2) decay length spectroscopy \cite{Stroscio_PRL_1986}. In the first technique, impurity scattering of quasiparticles generates interference patterns with characteristic dispersive wave vectors $\vec{q}(\omega)$ that can be inverted to reconstruct the band structure. Since $\vec{q}$ is the momentum transfer, QPI imaging resolves only $\textit{relative}$ momentum coordinates between two states. In the second technique, the $\textit{absolute}$, in-plane momentum  $k_{||}$ of quasiparticles can be extracted from the decay of their tunneling current with increasing sample-tip separation. By combining the two momentum-resolved techniques, we discover a $\Gamma$ electron pocket 75 meV above $E_F$. Our density functional theory (DFT) calculations reproduce the presence of empty states at $\Gamma$, and furthermore explain how their energies are tuned by the Se height $h_{\textnormal{Se}}$.

We grew films of FeSe on Nb-doped SrTiO$_3$(001) (0.5\%) via molecular beam epitaxy (MBE). The substrates were pretreated with deionized water for 90 min at 80$^\circ$C, followed by an O$_2$-anneal for 3 h at 1000$^\circ$C. We then transferred the substrates into our MBE chamber (base pressure 1$\times$10$^{-10}$ Torr) and degassed them at 670$^\circ$C. We deposited FeSe by co-evaporating Fe (99.995\%) and Se (99.999\%) with a molar flux ratio of 1:6 and substrate temperature 520$^\circ$C.  Afterwards, we typically annealed the samples for an additional 2 h between 500--600$^\circ$C before transferring them through ultra-high vacuum to a home-built STM for imaging at $\sim$4.3 K.

Figure~\ref{Fig1}(a) shows a typical film topography, with regions of bare SrTiO$_3$ and 1UC or 2UC of FeSe. We discriminate these regions based on their terrace heights. From the line cut in Fig.~\ref{Fig1}(b), we observe a 3UC SrTiO$_3$ step to be $1.19\pm0.05$ nm (bulk $c$-axis lattice constant is 0.3905 nm \cite{Schmidbauer2012_ACSB_2012}), the 1--2UC FeSe step to be $0.57\pm0.05$ nm, and the bare SrTiO$_3$--1UC FeSe step to be $0.34\pm0.02$ nm (all measured at 4 V sample-tip bias). We will hereafter focus on the 1UC FeSe terraces. Figure~\ref{Fig1}(c) presents an atomically-resolved topography of 1UC FeSe, with lattice constant $a$ = 3.9 \AA. Each bright spot corresponds to a surface Se atom in a Se-Fe-Se triple layer. A representative $dI/dV$ spectrum on a clean area exhibits a gap of $\Delta$ = 14 meV (Fig.~\ref{Fig1}(d)), similar in magnitude to other reports of superconducting gaps in this material ~\cite{Liu_NatComm_2012, Zhang_PRB_2014}. We note appreciable spectral inhomogeneity in 1UC FeSe/SrTiO$_3$, but further study is needed to quantify its correlation with substrate disorder.

To image QPI, we acquired conductance maps $g(\vec{r}, \omega) = dI/dV(\vec{r}, eV)$ over flat regions of 1UC FeSe/SrTiO$_3$ with moderate concentrations of as-grown defects (Fig.~\ref{Fig2}(a)). Several energy maps of one representative region are presented in Figs.~\ref{Fig2}(b-e), displaying clearly dispersive interference patterns. To identify the momentum-space origin of the scattered quasiparticles, we compared the Fourier transform amplitudes $\left|g(\vec{q}, \omega)\right|$ to simulated autocorrelations of the spectral function $A(\vec{k}, \omega) = -\frac{1}{\pi}\sum_{\alpha}\textnormal{Im}\lbrack G_{\alpha}(\vec{k}, \omega)\rbrack$ ~\cite{Chatterjee_PRL_2006}. For simplicity, we used the bare Green's function $G_{\alpha}^{-1}(\vec{k}, \omega) = \omega + i\delta - \varepsilon_{\alpha}(\vec{k})$, with parabolic bands $\varepsilon_{\alpha}(\vec{k})$ and broadening $\delta$ = 5 meV. The main result is presented in Figs.~\ref{Fig3}(a-i), which compare $\left|g(\vec{q}, \omega)\right|$ to theoretical predictions for three representative energies. We discuss each in turn:

\begin{figure*}[tb]
\includegraphics[scale=1]{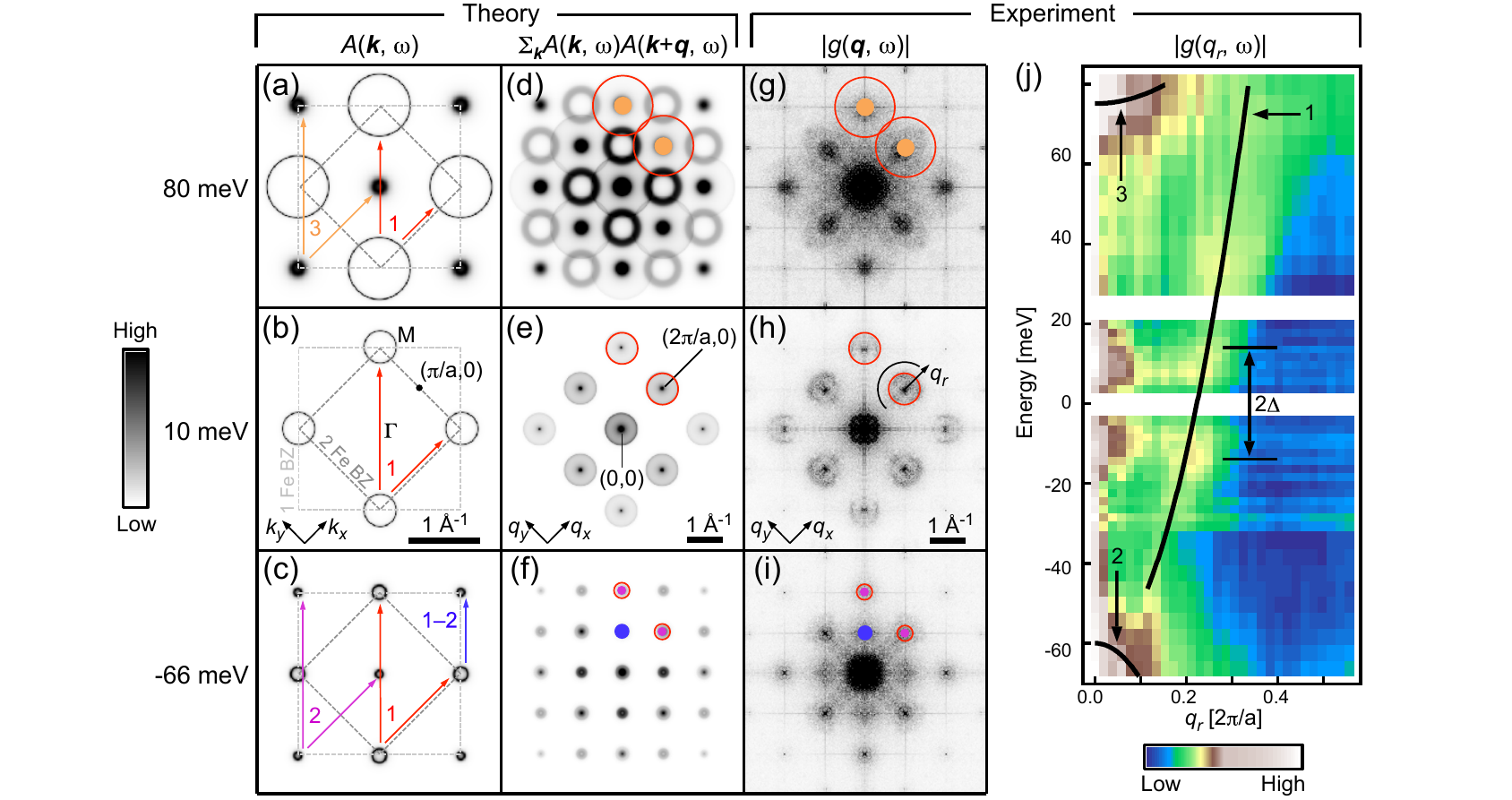}
\caption{(color online) Quasiparticle interference imaging, momentum transfer ($\vec{q}$) space. (a-f) Theoretical simulations, $A(\vec{k}, \omega)$ and its autocorrelation, for three representative energies. (g-i) Fourier transform amplitudes $\left|g(\vec{q}, \omega)\right|$ of conductance maps (four-fold symmetrized for increased signal). (j) Azimuthally-averaged intensity plot of $\left|g(q_r, \omega)\right|$, where $q_r$ is measured relative to $\vec{G} = (2\pi/a, 0)$. The superconducting gap is marked by $2\Delta$.}
\label{Fig3}
\end{figure*}

$\omega=10$ meV, Figs.~\ref{Fig3}(b,e,h): Close to $E_F$, we observe 9 ring-like intensities in $\left|g(\vec{q}, \omega)\right|$, centered about reciprocal lattice vectors $\vec{G}$ = (0, 0), ($\pm2\pi$/a, 0), (0, $\pm2\pi$/a), and ($\pm2\pi$/a, $\pm2\pi$/a). These intensities arise from scattering, modulo $\vec{G}$, within electron Fermi pockets at the zone corner M (labeled 1 in Fig.~\ref{Fig3}) \cite{Liu_NatComm_2012}.

$\omega=-66$ meV, Figs.~\ref{Fig3}(c,f,i): Sufficiently below $E_F$, we observe additional scattering channels pointing to the emergence of the $\Gamma$ hole pocket seen by angle-resolved photoemission spectroscopy (ARPES) \cite{Liu_NatComm_2012}. Intrapocket scattering between $\Gamma$ pockets is labeled 2 in Fig.~\ref{Fig3}, while interpocket scattering between $\Gamma$ and M pockets is labeled 1--2 in Fig.~\ref{Fig3}.

$\omega=80$ meV, Figs.~\ref{Fig3}(a,d,g): Above $E_F$, we discover a third pocket. Intrapocket scattering (labeled 3 in Fig.~\ref{Fig3}) is clearly resolved in $\left|g(\vec{q}, \omega)\right|$, but interpocket scattering with the M electron pockets (expected intensity at $(\pi/a, \pi/a)$ modulo $\vec{G}$) appears to be suppressed. In general, the autocorrelation of $A(\vec{k}, \omega)$ yields the set of all possible scattering channels, but more complex theories that encode spin \cite{Lee_PRB_2009} or orbital \cite{Zeljkovic_NatPhys_2014} selectivity in the scattering $T$-matrix are needed to explain their relative intensities. In this case, the empirical suppression of $\Gamma$--M scattering leaves some ambiguity as to the absolute momentum ($\vec{k}$) location of the new pocket.

To visualize the full QPI evolution, Fig.~\ref{Fig3}(j) shows an azimuthally-averaged intensity plot of $\left|g(q_r, \omega)\right|$, where $q_r$ is measured relative to $\vec{G} = (2\pi/a, 0)$ as shown in Fig.~\ref{Fig3}(h). In total, we observe three dispersing branches: two electron-like (labeled 1 and 3) and one hole-like (labeled 2). Branches 1 and 2 correspond to a M electron pocket and a $\Gamma$ hole pocket, while branch 3 awaits further identification. A parabolic fit to branch 1 over the given energy range in Fig.~\ref{Fig3}(j) yields an effective mass enhancement $m^*/m = 2.0 \pm 0.1$ and a carrier concentration of 0.08 $e^{-}$ per Fe from a Luttinger count, assuming a degenerate pocket \cite{Liu_NatComm_2012, SM}.

\begin{figure}[b]
\includegraphics[scale=1]{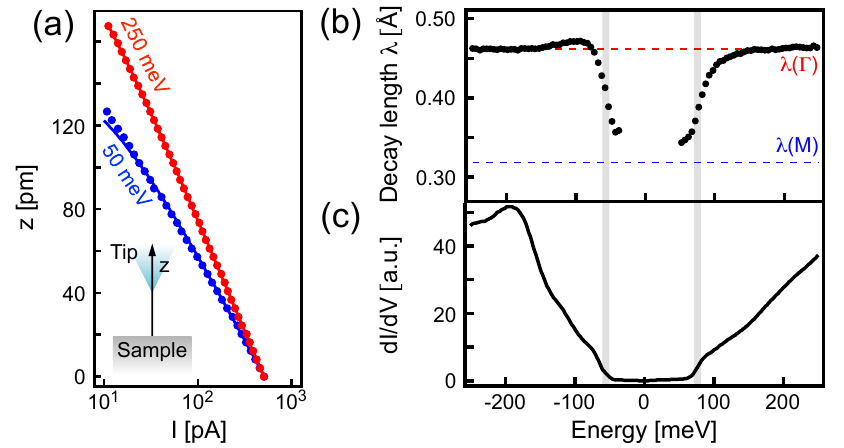}
\caption{(color online). (a,b) Energy dependent decay length $\lambda(\omega)$, extracted from exponential fits to the tunneling current as the tip is retracted from the sample at a fixed bias (inset schematic). Fits were performed in the current range [10 pA, 500 pA], two of which (250 meV and 50 meV) are shown in (a). Dashed horizontal lines indicate calculated values of $\lambda$ at the $\Gamma$ and M. (c) $dI/dV$ spectrum. $V_{\textnormal{rms}}$ = 2.8 mV. Vertical lines mark extrema in the numerical derivatives of $\lambda(\omega)$ and $dI/dV$.}
\label{Fig4}
\end{figure}

To determine the absolute momentum $\vec{k}$ of QPI branch 3, a complimentary momentum-resolved STM technique is needed. Here we utilize decay length spectroscopy \cite{Stroscio_PRL_1986, Zhang_arXiv_2014, Zhang_NatPhys_2008}, a general tool which allows the full reconstruction of $\vec{k}$-space band structure from STM. Tersoff and Hamman \cite{Tersoff_PRL_1983} showed that a sample state of in-plane momentum $\vec{k_{||}}$
has density which decays towards the vacuum with length $\lambda$ given by
\begin{equation}\label{Eq:lambda}
\frac{1}{(2\lambda)^2} = \frac{2m\Phi}{\hbar^2} + k_{||}^2,
\end{equation}
where $\Phi$ is the average of the sample and tip work functions. Figures~\ref{Fig4}(a,b) show the energy dependent decay length $\lambda(\omega)$, extracted from exponential fits to the tunneling current as the sample-tip distance is increased at a fixed bias. Near $E_F$, the sample states have large momentum near M and smaller decay length. Below $E_F$, a steep increase in $\lambda(\omega)$ accompanies the onset of a hole pocket at $\Gamma$, as states with low momentum become available for tunneling. The fact that a similar rise in $\lambda(\omega)$ occurs above $E_F$ indicates that branch 3 in Fig.~\ref{Fig3}(j) is also located at $\Gamma$. If we interpret the large-$\left|\omega\right|$ value of $\lambda = 0.462 \pm 0.001$ \AA\ as arising from states with $\vec{k}\approx 0$, we find $\Phi = 4.46 \pm 0.03$ eV from Eq.~(\ref{Eq:lambda}), then we can compute the expected $\lambda(\omega)=0.318 \pm 0.001$ \AA\ for energies where the only states come from momenta near M. Indeed, the measured $\lambda(\omega)$ at small $\left|\omega\right|$ closely matches the expected value of $\lambda(\left|\vec{k}\right|=\sqrt{2}\pi/a)$. Step-like features associated with the onsets of these pockets are also detected with $dI/dV$ spectroscopy (Fig.~\ref{Fig4}(c)). From extrema in the numerical derivative $d^2I/dV^2$, which closely match those of $d\lambda/d\omega$ (vertical shaded guides in Figs.~\ref{Fig4}(b,c)), the band edges of the $\Gamma$ hole and electron pockets are -65 meV and 75 meV.

\begin{figure}[tb]
\includegraphics[scale=1]{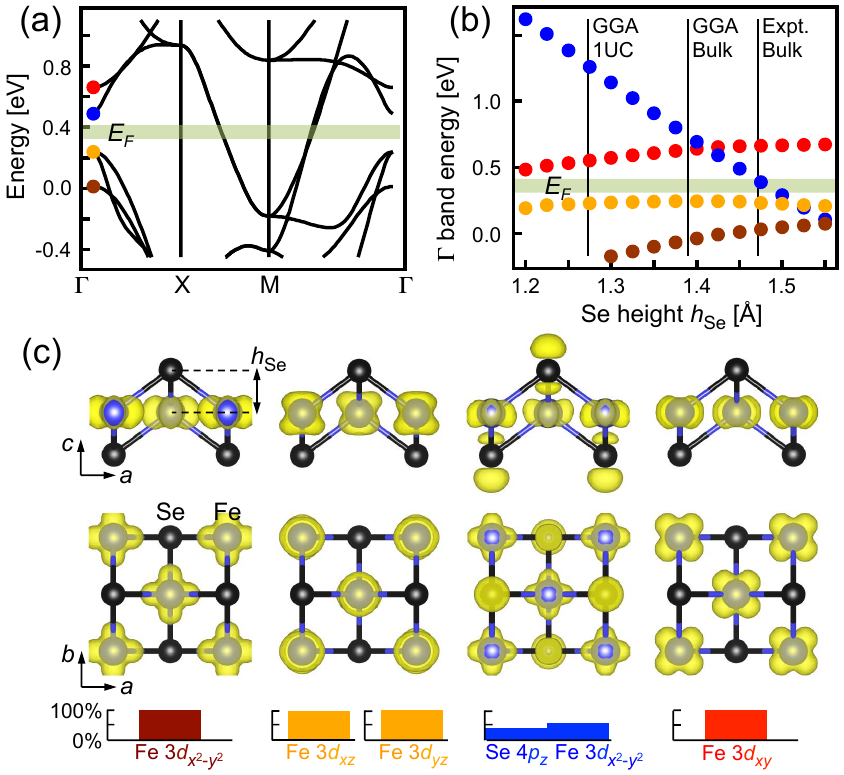}
\caption{(color online) (a) Band structure of free-standing single-unit-cell (1UC) FeSe, calculated in the generalized gradient approximation (GGA). Structural parameters: lattice constant $a$ = 3.90 \AA, Se height $h_{\textnormal{Se}}$ = 1.45 \AA. (b) Energies of the five $\Gamma$ bands shown in (a) vs.\ $h_{\textnormal{Se}}$ (the band represented by orange is degenerate). The GGA values of $h_{\textnormal{Se}}$ for 1UC FeSe ($a$ = 3.90 \AA\ fixed) and bulk FeSe ($a$ = 3.68 \AA\ relaxed) are marked, as well as the experimental value for bulk FeSe. The Fermi energy $E_F$ expected from electron doping is marked in (a,b). (c) Charge density isosurfaces (yellow) at $\vec{k} = 0$ for the five $\Gamma$ bands, shown in two perspectives. The histograms depict the orbital compositions.}
\label{Fig5}
\end{figure}

A consistent band structure for 1UC FeSe/SrTiO$_3$ is now established, comprising M electron pockets spanning $E_F$ and $\Gamma$ hole and electron pockets lying below and above $E_F$. For further insight, we use DFT to compute the band structure of free-standing 1UC FeSe via the generalized gradient approximation (GGA) \cite{Perdew_PRL_1996} and projector augmented wave (PAW) method as implemented in the Vienna Ab-Initio Simulation Package (VASP) \cite{Kresse_PRB_1996, Kresse_CMS_1996}.  We use a BZ sampling of 9$\times$9$\times$1 and an energy cutoff of 450 eV. We apply Methfessel-Paxton smearing \cite{Methfessel_PRB_1989} with $\sigma$ = 0.1 eV. Figure~\ref{Fig5}(a) shows the calculated bands with structural parameters $a$ = 3.90 \AA, $h_{\textnormal{Se}}$ = 1.45 \AA. Due to electron doping, $E_F$ should be adjusted to intersect only the M pockets. Typical band renormalization factors range from 4--5 in 1UC FeSe/SrTiO$_3$ \cite{Peng_NatComm_2014}, but for the qualitative discussion that follows, we do not rescale the bands.

Experimentally, $h_{\textnormal{Se}}$ is unknown. Simulations show that the binding geometry of 1UC FeSe/SrTiO$_3$ varies with TiO$_2$ oxygen deficiency, which creates electropositive sites that distort Se positions \cite{Bang_PRB_2013}. Without microscopic knowledge of the buried interface, we calculate band structures for a range of $h_{\textnormal{Se}}$ values and track the energies of the $\Gamma$ bands (Fig.~\ref{Fig5}(b)). While all bands shift slightly, the lowest-lying $\Gamma$ electron pocket in Fig.~\ref{Fig5}(a) undergoes a pronounced monotonic decrease in energy with increasing $h_{\textnormal{Se}}$. Figure~\ref{Fig5}(c) shows the charge density isosurfaces at $\vec{k} = 0$ and orbital compositions for each band. Only the lowest-lying $\Gamma$ electron pocket carries significant Se $4p$ character in addition to Fe $3d$ character, so it is most affected by the Fe-Se distances. The charge density plot suggests an antibonding configuration of Fe $3d_{x^{2}-y^{2}}$ and Se 4$p_z$ orbitals, which explains the increase in pocket energy with greater overlap of Fe and Se states. Our calculation reveals a crucial connection between $h_{\textnormal{Se}}$ and empty electronic states.

Previous reports have predicted that Se/Te heights tune the Fe exchange constants in iron chalcogenides and hence the magnetic order \cite{Moon_PRL_2010}, which is oddly absent in FeSe \cite{Medvedev_NatMat_2009} and unknown in 1UC FeSe/SrTiO$_3$. Here, we discuss another implication of $h_{\textnormal{Se}}$.  As seen in Fig.~\ref{Fig5}(b), the $\Gamma$ electron and hole pockets cross at large values of $h_{\textnormal{Se}}$. Recently, Wu \textit{et al.} have proposed that nontrivial $\mathbb{Z}_2$ topology may be realized in 1UC FeTe$_{1-x}$Se$_x$ \cite{Wu_arXiv_2014}. In particular, when the gap $\Delta_n$ between the $\Gamma$ electron and hole pockets falls below 80 meV, spin-orbit coupling can invert the bands.  We measure $\Delta_n$ to be 140 meV from Fig.~\ref{Fig4}(c); thus, 1UC FeSe/SrTiO$_3$ could possibly lie in proximity to a topological phase transition.

In summary, we have quantified both the filled and empty state band structure of 1UC FeSe/SrTiO$_3$, and discovered a new $\Gamma$-centered pocket emerging around 75 meV above the Fermi level. Our work has several important implications, both for superconductivity and for predicted topological order in FeSe/SrTiO$_3$. First, the new $\Gamma$ band will serve as an essential input for revised FRG calculations of the effective low-energy pairing interaction \cite{Xiang_PRB_2012}. Second, the modest 140 meV gap we measured between filled and empty Gamma bands gives hope that inversion of these bands may be achievable, and may lead to a predicted topological phase \cite{Wu_arXiv_2014}. Finally, our work introduces decay length spectroscopy as a general and complementary technique to QPI imaging, to map the absolute momentum-resolved electronic band structure of filled and empty states using STM. We suggest the use of these techniques in concert to track the $\Gamma$ pocket energies in future strain engineering experiments with FeSe.

\begin{acknowledgments}
We thank P. J. Hirschfeld, I. I. Mazin, Subir Sachdev, B. I. Halperin, and Sinisa Coh for useful conversations. This work was supported by the National Science Foundation under grants DMR-0847433 and DMR-1231319 (STC Center for Integrated Quantum Materials), and the Gordon and Betty Moore Foundation’s EPiQS Initiative through grant GBMF4536.  Computations were run on the Odyssey cluster supported by the FAS Division of Science, Research Computing Group at Harvard University. D. H. acknowledges support from an NSERC PGS-D fellowship. C. L. S. acknowledges support from the Lawrence Golub fellowship at Harvard University. S. F. and E. K. acknowledge support by ARO-MURI W911NF-14-1-0247. J. E. H. acknowledges support from the Canadian Institute for Advanced Research.
\end{acknowledgments}

\newpage

\setcounter{figure}{0}

\renewcommand{\thefigure}{S\arabic{figure}}	
\setcounter{figure}{0}
\renewcommand{\thetable}{S\arabic{table}}	
\setcounter{table}{0}
\renewcommand{\theequation}{S\arabic{equation}}	
\setcounter{equation}{0}

\onecolumngrid

{\Large \textbf{Supplementary Material for:}}
\begin{center}
{\large \textbf{Revealing the Empty-State Electronic Structure of Single-Unit-Cell FeSe/SrTiO$_{3}$}}

Dennis Huang, Can-Li Song, Tatiana A. Webb, Shiang Fang, Cui-Zu Chang, Jagadeesh S. Moodera, \\Efthimios Kaxiras, Jennifer E. Hoffman
\end{center}

\setcounter{figure}{0}
\setcounter{equation}{0}
\setcounter{table}{0}
\makeatletter
\renewcommand{\thefigure}{S\@arabic\c@figure}
\renewcommand{\theequation}{S\@arabic\c@equation}
\renewcommand{\thetable}{S\@arabic\c@table}
\vspace{2mm}

\begin{figure}[!h]
\includegraphics[scale=1]{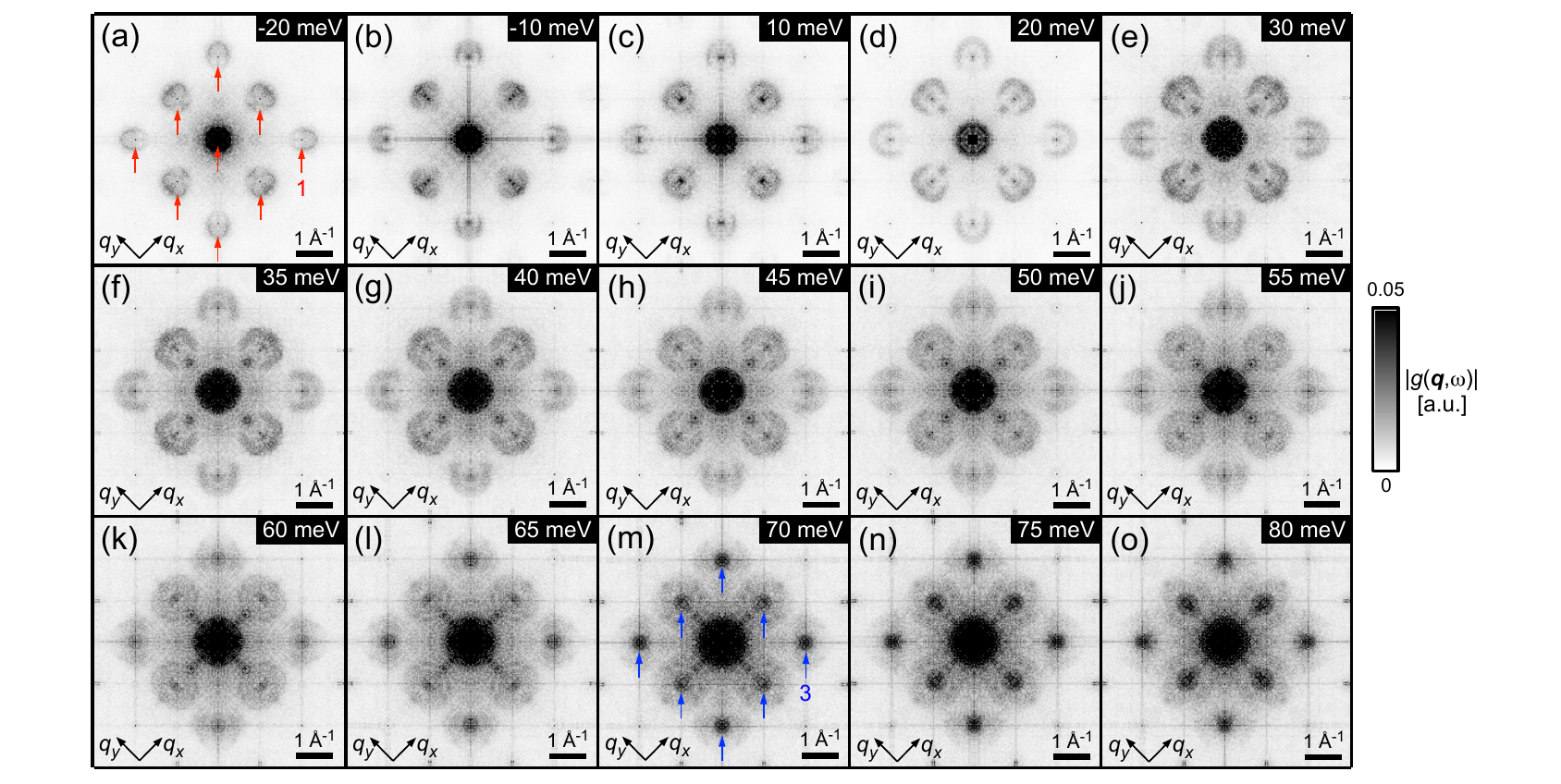}
\caption{Quasiparticle interference (QPI) imaging, momentum-transfer ($\vec{q}$) space. (a-o) Fourier transform amplitudes $|g(\vec{q},\omega)|$ of conductance maps used in Fig. 3(j) of the main text. All images are displayed with the same color scale range. The red and blue arrows denote QPI branches 1 and 3 as discussed in the main text.}
\label{FigS1}
\end{figure}

Figure~\ref{FigS1} presents the individual energy layers of $|g(\vec{q},\omega)|$ that constitute the high-energy portion of Fig. 3(j) of the main text. There is one set of rings at -20 meV that grows with increasing energy (branch 1 in Fig. 3(j), red in Fig.~\ref{FigS1}), and another set of higher intensity rings that first appears around 70 meV (branch 3 in Fig. 3(j), blue in Fig.~\ref{FigS1}). 

\begin{figure}[!h]
\includegraphics[scale=1]{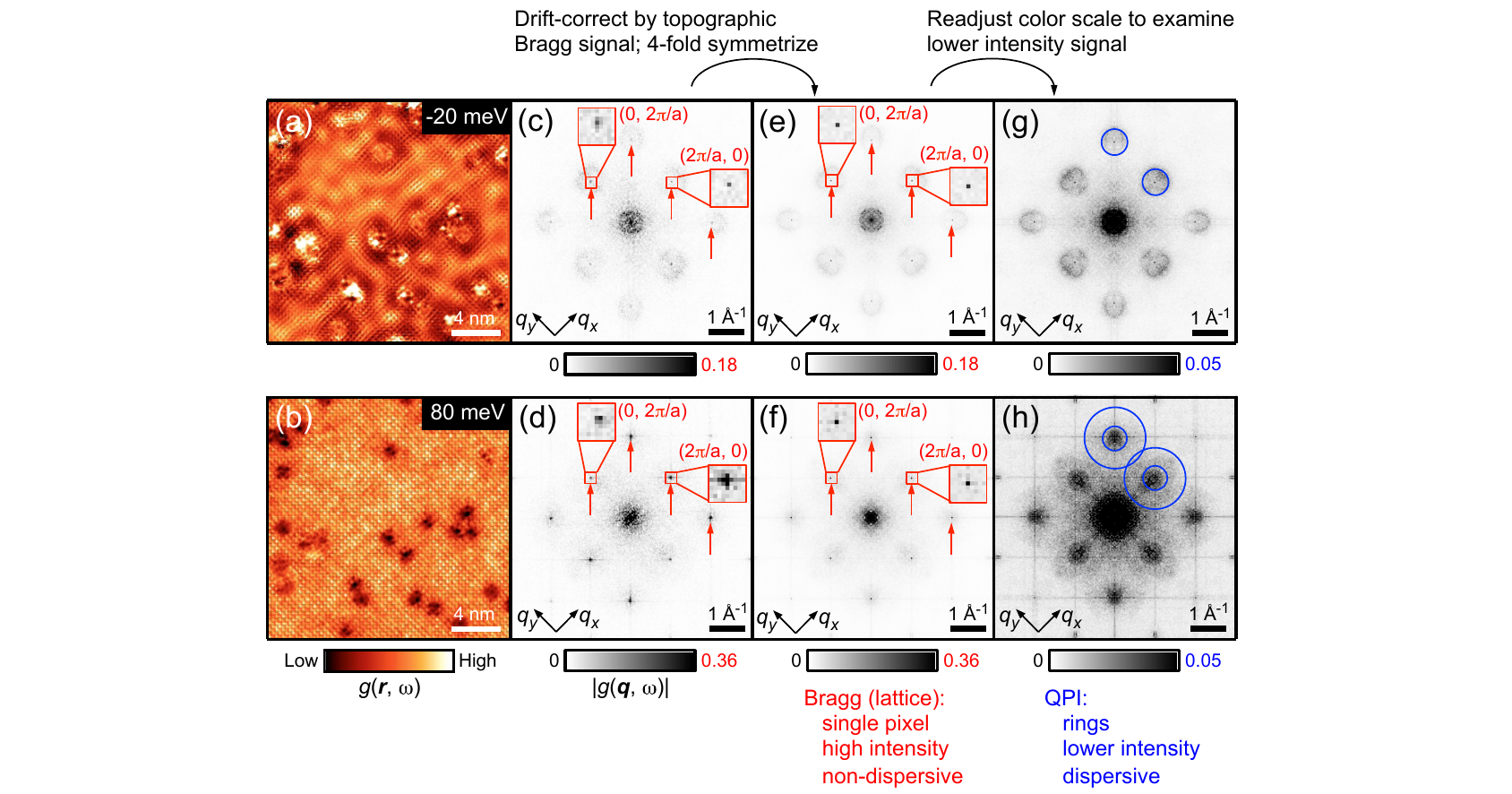}
\caption{Distinct observations of Bragg signals \textit{at} the reciprocal lattice vectors and QPI signals \textit{around} the reciprocal lattice vectors. (a,b) Conductance maps at two energy layers, -20 meV and 80 meV, along with their Fourier transform amplitudes $|g(\vec{q},\omega)|$ in (c,d). (e,f) After applying a drift-correction algorithm using parameters determined entirely from the simultaneous topographic map, the Bragg peaks appear as a non-dispersive, single-pixel entities. (g,h) By examining lower intensities of (e,f), QPI signals manifest as dispersive rings with larger $\vec{q}$-space radius.}
\label{FigS2}
\end{figure}

In general, near band edge energies, both quasiparticle interference (QPI) and Bragg (lattice) signals may have weight near the reciprocal lattice vectors. Here, we illustrate in Fig.~\ref{FigS2} that our data exhibit three characteristic distinctions between the Bragg and QPI signals: (1) $\vec{q}$-space extent, (2) intensity, and (3) dispersion.  The Bragg peaks are only a few pixels wide in the raw data; they have large intensity; and they appear in the same $\vec{q}$-space location across all energies.  Furthermore, they collapse onto a single pixel after applying drift-correction according to the \textit{topographic} lattice simultaneously acquired (see Refs.~\cite{Lawler_Nat_2010, Zeljkovic_Nat_Mat_2012} for extensive details). On the other hand, the QPI signals possess a larger radius in $\vec{q}$-space; they appear at lower intensities; and they disperse/appear/disappear with energy. The QPI signal that first appears around 70 meV, corresponding to the emergence of an empty-state band, cannot be explained by an abnormally large smearing of the Bragg signal.

\begin{figure}[!h]
\includegraphics[scale=1]{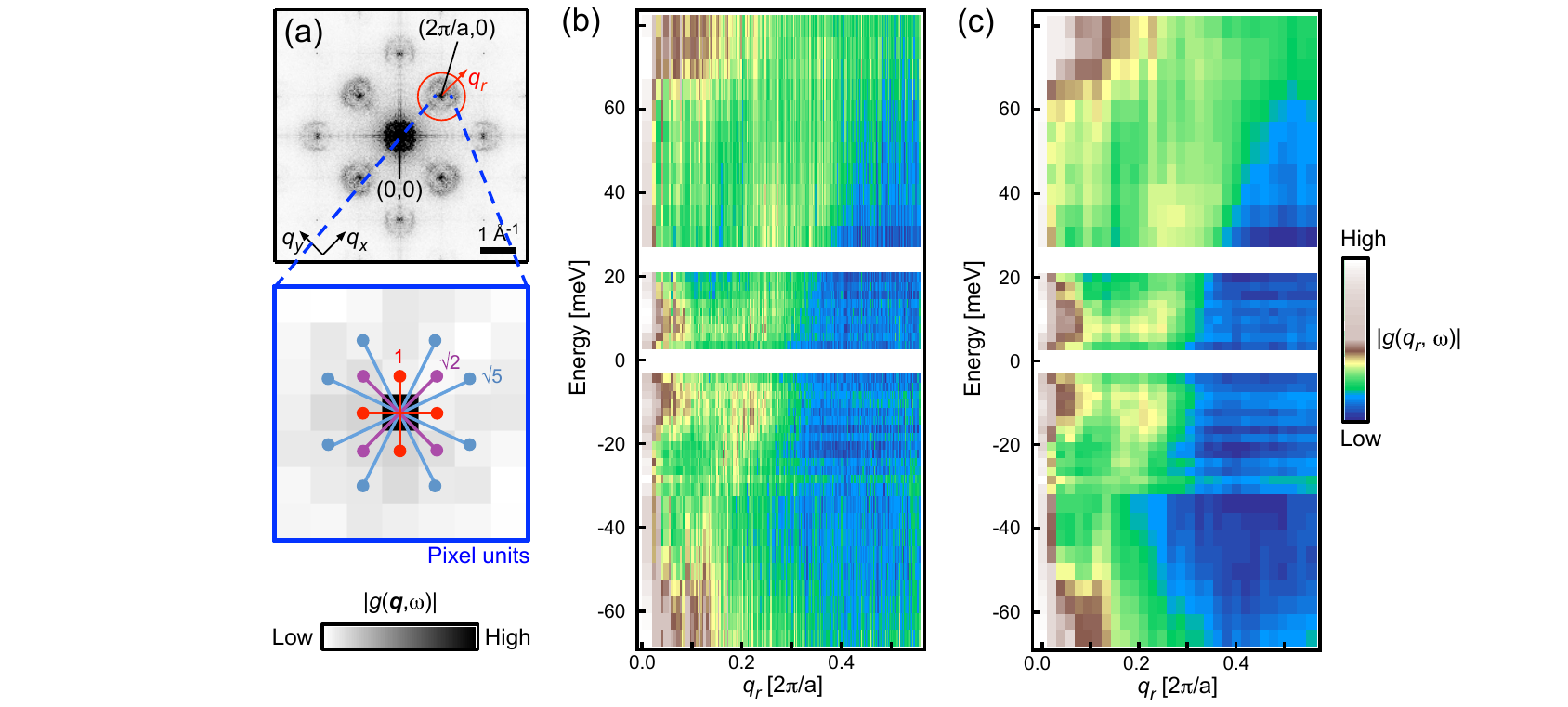}
\caption{Azimuthal averaging procedure used for visualizing QPI dispersion. (a,b) First, every pixel of $|g(\vec{q},\omega)|$ near $\vec{G} = (2\pi/a,0)$ is averaged with those having identical distance values $q_r$ to $\vec{G}$, then displayed as a function of $q_r$. (c) Second, each discrete $q_r$ value is binned with its nearest integer pixel value to produce the final plot shown in Fig. 3(j).}
\label{FigS3}
\end{figure}

Figure~\ref{FigS3} illustrates the azimuthal averaging procedure used to produce Fig.\ 3(j). Within a large radius of $\vec{G} = (2\pi/a,0)$ encompassing the QPI signal, every pixel of $|g(\vec{q},\omega)|$ is averaged with those having identical distance values $q_r$ to $\vec{G}$, then displayed as a function of $q_r$. The resulting plot makes full use of our $\vec{q}$-space pixel resolution, but the discrete $q_r$ values are spaced unevenly (Figs.~\ref{FigS3}(a,b)). We bin each discrete $q_r$ value with its nearest integer pixel value to produce the final plot in Fig.~\ref{FigS3}(c). We additionally note that the first pixel column in  Figs.~\ref{FigS3}(b,c) is nearly saturated because the Bragg peak intensity is so much greater than the QPI signal.

\begin{figure}[!h]
\includegraphics[scale=1]{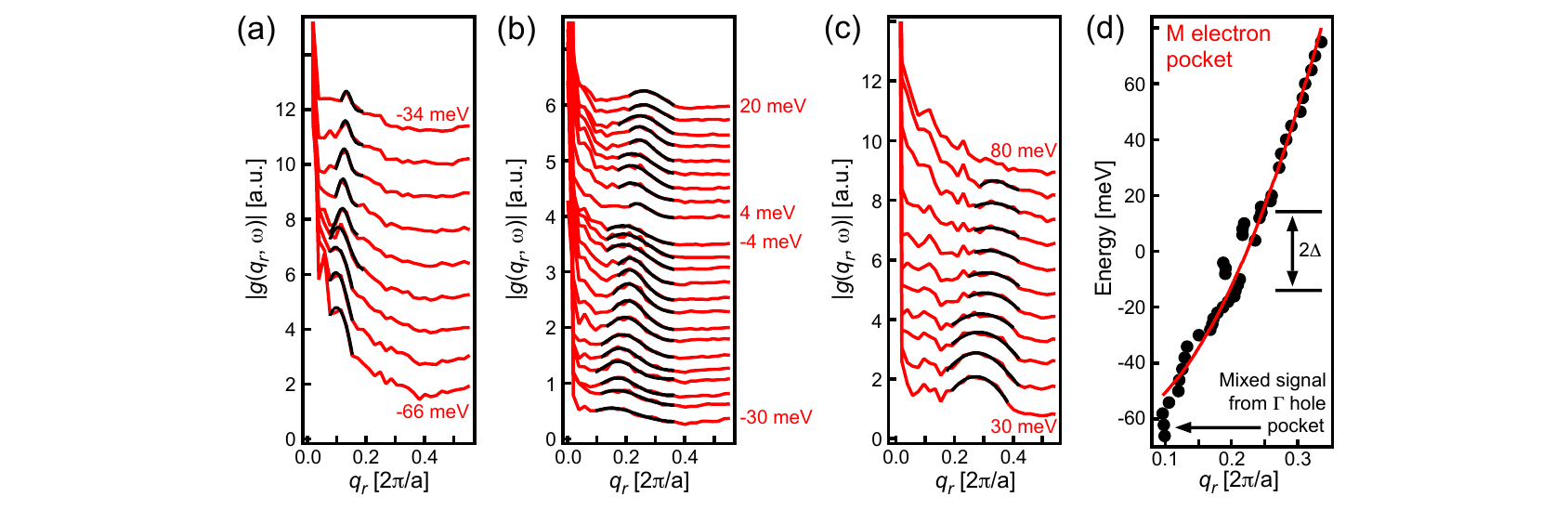}
\caption{(a-c) Azimuthally-averaged line cuts of $|g(q_r,\omega)|$, offset in evenly-spaced energies. $q_r$ is measured relative to $\vec{G} = (2\pi/a,0)$. Dispersing peaks from M electron pocket scattering (branch 1 in the main text) are fit to Gaussians (black line overlays), and the extracted peak positions are fit to a parabola in (d). The superconducting gap is marked by 2$\Delta$.}
\label{FigS4}
\end{figure}

The three bands observed in this work (M electron pocket, $\Gamma$ hole pocket, $\Gamma$ electron pocket) are quantified by a combination of three techniques: (1) QPI $|g(q_r,\omega)|$, (2) decay length spectroscopy $d\lambda/d\omega$, and (3) $d^2I/dV^2$. Figures~\ref{FigS4}(a-c) show constant energy cuts of $|g(q_r,\omega)|$, with prominent dispersing peaks due to M electron pocket scattering. These dispersing peak positions are fit to a parabola (Fig.~\ref{FigS4}(d)), which then serves as the guide overlaid in Fig. 3(j) of the main text and labeled ``branch 1''. Slight deviations from the parabolic fit are present within the superconducting gap energy 2$\Delta$.

\begin{figure}[!h]
\includegraphics[scale=1]{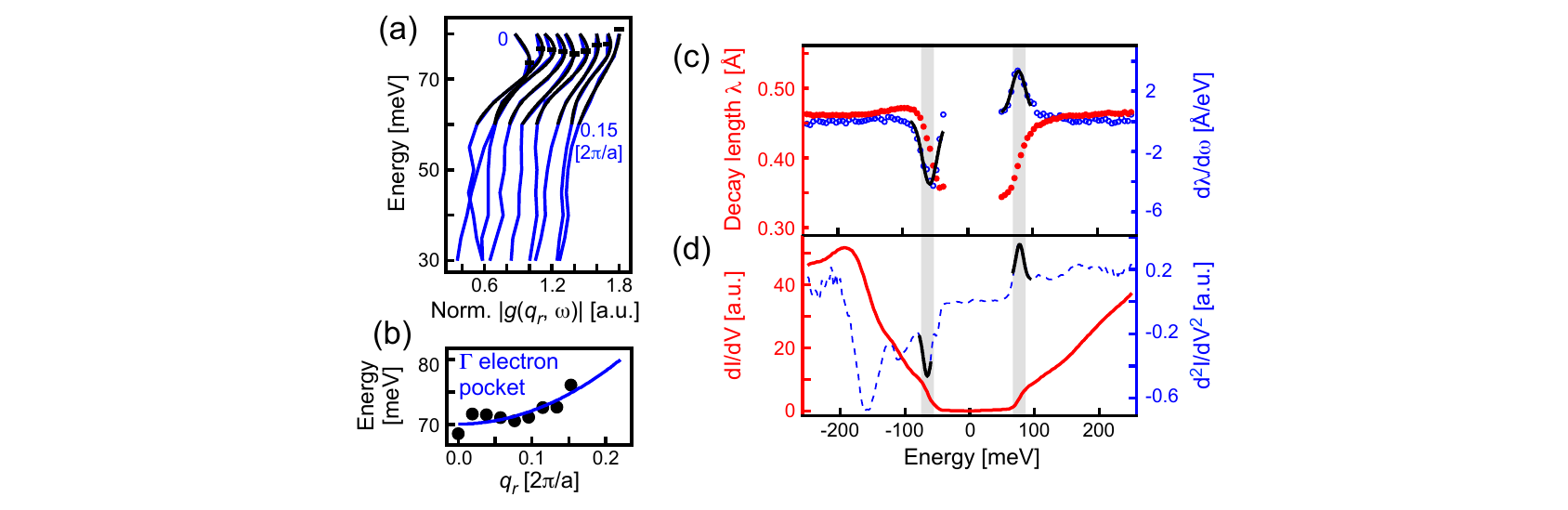}
\caption{(a) Constant-$q_r$ line cuts of $|g(q_r,\omega)|$, offset in evenly-spaced $q_r$ values and normalized by their maximum amplitude for improved visualization. Dispersing peaks from $\Gamma$ electron pocket scattering (branch 3 in the main text) are fit to Gaussians (black line overlays with horizontal bars indicating peak positions), and the resulting peak positions are fit to a parabola in (b). (c) Energy dependent decay length $\lambda(\omega)$ (red circles), repeated from Fig. 4(b) of the main text, along with its numerical derivative (blue circles). (d) $dI/dV$ spectrum (red solid line), repeated from Fig. 4(c) of the main text, along with its numerical derivative (blue dashed lines). Gaussian peak fits for band edge extraction are overlaid (black solid lines), and vertical shaded bars mark their peak positions.}
\label{FigS5}
\end{figure}

QPI branches 2 and 3 are visible in Figs.~\ref{FigS4}(a,c) as emerging peaks centered about $q_r=0$ that grow in amplitude and width away from the Fermi energy $E_F$. The band edge for branch 2 is difficut to fit, along either the energy or $q_r$ axes, because its signal overlaps with that of branch 1. However, its hole-like dispersion is evident when one tracks the tail end of the dispersing peaks (green pixels in Fig.~\ref{FigS3}(c) within $q_r$ $\sim$ 0.2--0.3 [2$\pi$/a]) and observes a change in dispersion direction near -60 meV. (A similar kink in the dispersion is visible at high energies, corresponding to the onset of branch 3). We quantify branch 2 by tunneling decay length and $dI/dV$ measurements. Details are discussed in the main text, but here in Fig.~\ref{FigS5}(c,d) the numerical derivatives $d\lambda/d\omega$ and $d^2I/dV^2$ and their peak fits are explicitly shown. We overlay a guide for branch 2 in Fig. 3(j) with its band edge informed by these two measurements (-59 $\pm$ 5 meV and -65 $\pm$ 3 meV respectively) and its dispersion informed by angle-resolved photoemission spectroscopy (ARPES) measurements of the same pocket \cite{Liu_NatComm_2012}.

To capture the band edge of branch 3, we take constant-$q_r$ cuts of $|g(q_r,\omega)|$, and fit the observed peak locations to a parabolic dispersion (Figs.~\ref{FigS5}(a,b)). The resulting fit serves as the guide for branch 3 overlaid in Fig. 3(j). The position of this line is further confirmed by the positive energy extrema of $d\lambda/d\omega$ and $d^2I/dV^2$, which yield 78 $\pm$ 5 meV and 77 $\pm$ 3 meV respectively.

A summary of the band parameters extracted from these quantitative analyses is given in Table~\ref{TableS1}.

\begingroup
\begin{table}[!h]
\center
\scalebox{1}{
\setlength{\tabcolsep}{10pt}
\begin{tabular}{*7c}
\hline\hline
Main text & SM figure & Method & \multicolumn{2}{c}{\underline{M electron pocket}} & \underline{$\Gamma$ hole pocket} & \underline{$\Gamma$ electron pocket}  \\ 
figure & & & $k_F$ [$\pi/a$] & $m^*/m$ & $\varepsilon_0$ [meV] & $\varepsilon_0$ [meV] \\
\hline
3(j) & \ref{FigS4}, \ref{FigS5}(a,b) & $|g(q_{r}, \omega)|$ & $0.22 \pm 0.01$ & $2.0 \pm 0.1$ & (-60) & $75 \pm 3$  \\
4(b) & \ref{FigS5}(c) & $d\lambda/d\omega$ & -- & -- & -$59 \pm 5$ & $78 \pm 5$ \\
4(c) & \ref{FigS5}(d) & $d^2I/dV^2$ & -- & -- & -$65 \pm 3$ & $77 \pm 3$ \\
\hline\hline
\end{tabular}}
\caption{Summary of band parameters quantified by QPI imaging $|g(q_{r}, \omega)|$, decay length spectroscopy $d\lambda/d\omega$, and $d^2I/dV^2$. $k_F$ denotes the Fermi wave vector, $m^*$ is the effective mass, and $\varepsilon_0$ is the band edge.}
\label{TableS1}
\end{table}
\endgroup

\begin{figure}[!h]
\includegraphics[scale=1]{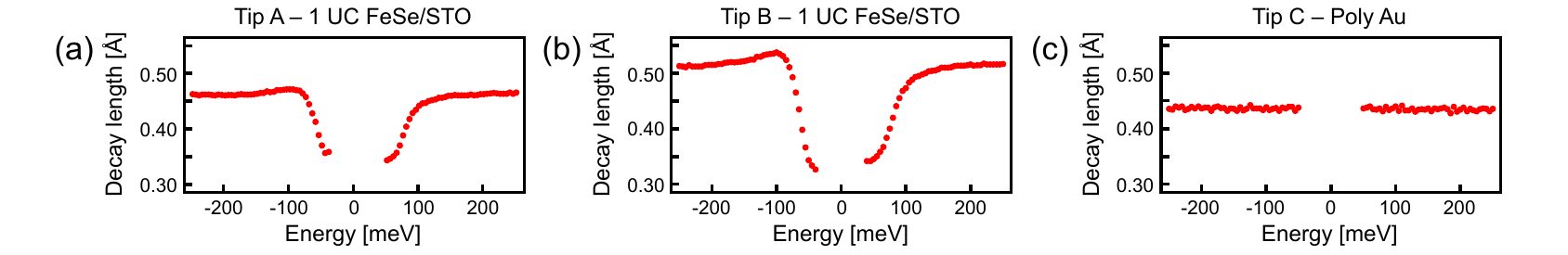}
\caption{(a,b) Energy dependent decay length $\lambda(\omega)$ of single-unit-cell (1UC) FeSe/SrTiO$_3$, taken with a PtIr tip with two different microscopic terminations (called A and B). (c) $\lambda(\omega)$ acquired on polycrystalline Au.}
\label{FigS6}
\end{figure}

\newpage Finally, we present additional decay length spectroscopy measurements. Figures~\ref{FigS6}(a,b) compare $\lambda(\omega)$ measurements of single-unit-cell FeSe/SrTiO$_3$ taken with a PtIr tip with two different microscopic terminations. While the absolute values of $\lambda(\omega)$ vary slightly due to differences in tip work function, the steep rises corresponding to the onset of $\Gamma$-centered pockets occur at the same energies. Figure~\ref{FigS6}(c) shows a calibration measurement on polycrystalline Au exhibiting a flat $\lambda(\omega)$, consistent with previous measurements on Au(111) \cite{Zhang_NatPhys_2008}. 

\end{document}